\documentclass[%
 preprint,
superscriptaddress,
footinbib,
amsmath,amssymb,
prd,
]{revtex4-1}

\usepackage[english]{babel}
\usepackage[utf8x]{inputenc}
\usepackage[colorinlistoftodos]{todonotes}
\usepackage{enumerate}
\usepackage{graphicx}
\usepackage{dcolumn}
\usepackage{bm}

\usepackage{textcomp}
\usepackage[pagewise,columnwise]{lineno}

\begin{document}

\preprint{APS/123-QED}

\title{
On the relation between the soft and hard parts of the transverse momentum distribution}

\author{C. Pajares}
\email{pajares@fpaxp1.usc.es}
\affiliation{
Departamento de F\'isica de Part\'iculas and Instituto Galego de Física de Altas Enerxías, Universidad de Santiago de Compostela, E-15782 Santiago de Compostela, Espa\~na}

\author{J. E. Ram\'irez}
\email{jhony.ramirezcancino@viep.com.mx}
\affiliation{Centro de Agroecología,
Instituto de Ciencias,
Benemérita Universidad Autónoma de Puebla, Apartado Postal 165, 72000 Puebla, Puebla, M\'exico}

\begin{abstract}
Usually, the transverse momentum distribution is described by a sum of an exponential decay term plus a decreasing power like contribution representing the soft non-perturbative and hard perturbative QCD collisions, respectively.
In this paper, we derive an analytical function that can describe the whole transverse momentum spectrum. This is obtained using a $q$-Gaussian distribution to describe the string tension fluctuations. The parameter $q$ determines the departure of the thermal distribution as well as the minimum length that can be explored at high transverse momentum.
We also show that the ratio between both scales only depends on the $q$ parameter.
\end{abstract}
\maketitle


\section{Introduction}
Usually, the transverse momentum distribution (TMD) is described by a sum of an exponential decay term plus a decreasing power like contribution representing the soft non-perturbative and hard perturbative QCD collisions, respectively \cite{BYLINKIN201465, bylinkin}.
The presence of an exponential shape in the TMD of produced particles in collisions of small objects together with the approximate thermal abundances of the hadron yields constitutes an indicative sign of thermalization.
This thermalization however cannot be achieved under the usual mechanism, namely, final state interactions in the form of several secondary collisions.

In some models like the color string percolation model, this thermalization arise from the string tension fluctuations of strings or string clusters formed in the collisions.
The typical Gaussian form of the TMD of particles produced from the decay of strings or string clusters via the Schwinger mechanism is transformed into a thermal distribution assuming a Gaussian form of the string tension fluctuations \cite{BIALAS1999301, DIASDEDEUS2006455}. 
This is in line with the thermalization due to the Hawking-Unruh effect applied to hadronic collisions where the Schwinger mechanism is also working \cite{KHARZEEV2005316}.

The relation between non-perturbative and perturbative physics has been pointed out recently by means of the \emph{resurgent} extrapolations \cite{COSTIN2020135627}, where using analytical continuation and other techniques are able to join both domains under some requirements.

On the other hand, theoretical studies of quenches in entangled systems described by (1+1)-dimensional conformal theories of expanding quantum fields and strings, have shown that these systems behave as a generalized Gibbs ensemble with an effective temperature set by the energy cutoff for the ultraviolet modes \cite{Calabrese_2016, BERGES2018442}.
In a high energy collision, a hard interaction between two partons of the initial state of the colliding hadrons produces a rapid quench of the entangled partonic state, and thus the characteristic effective temperature -inferred from the TMD of the particles produced in the collisions- can depend on the energy scale of the hard process, which works as an ultraviolet cutoff of the modes resolved by the collision \cite{kharzeev, armesto2019, gotsman, tu,Ramos}.
This possibility has been recently studied in charged particles and Higgs boson production in pp and Pb-Pb collisions at very different energies and multiplicities, where was found that the ratio between the hard scale and the effective temperature is approximately 4 \cite{baker, feal, Bellwied_2018}.
This study was extensively performed including Xe-Xe collisions showing that the factor between both scales has to do with the normalized fluctuations of the hard scale.
In the case of nucleus-nucleus collisions due to final state interactions like jet quenching some departure of this behavior was observed \cite{feal2}.
It was pointed out the possibility of encoding in a single analytical function the whole transverse momentum distribution in such a way that the relation between both scales become quite clear.
This point is the main goal of the present study.
We show that using a Tsallis distribution instead of a Gaussian one to describe the string tension fluctuations we obtain a function which keeps the exponential decay function at low $p_T$ and shows a decreasing power like behavior at high $p_T$.
The ratio between the soft and hard scales depends only on $q$, whose value depend on the energy and multiplicity  of the considered process.

\section{Analytical universal function for the whole spectrum}

\begin{figure}[h]
\centering
\includegraphics[scale=1]{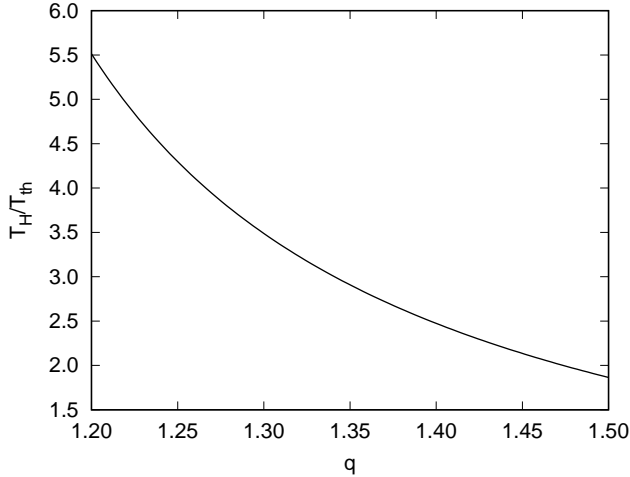}
\caption{Ratio $T_H/T_{th}$ as a function of $q$.}
\label{fig:ratio}
\end{figure}

Let us assume that the fluctuations of the tension in the decay of one string is a $q$-Gaussian distribution (instead of a Gaussian distribution) which for $1<q<3$ reads \cite{budini}
\begin{equation}
    N_q(x,\sigma)=\frac{\sqrt{q-1}\Gamma\left(\frac{1}{q-1}\right)}{\sigma \sqrt{2\pi} \Gamma\left(\frac{q-3}{2(1-q)}\right)} \left( 1+\frac{(q-1)x^2}{2\sigma^2}  \right)^\frac{1}{1-q}.
\end{equation}
In this way, the TMD is computed as the convolution of the Schwinger mechanism with the string tension fluctuations as follows
\begin{equation}
    \frac{dN}{dp_T^2}\sim \int_0^\infty \exp \left( -\frac{\pi p_T^2}{x^2}  \right) N_q(x,\sigma)dx.
\end{equation}
Introducing the variable
\begin{equation}
    \tau=\frac{2\sigma^2}{(q-1)x^2},
\end{equation}
we found
\begin{equation}
    \frac{dN}{dp_T^2}\sim \frac{\Gamma \left( \frac{1}{q-1} \right)}{\sqrt{\pi} \Gamma\left( \frac{q-3}{2(1-q)} \right)} \int_0^\infty \exp\left( -\frac{\pi p_T^2(q-1)}{2\sigma^2}\tau  \right) \tau^{\frac{1}{q-1}-\frac{3}{2}} (1+\tau)^{\frac{1}{1-q}} d\tau. \label{eq:conv}
\end{equation}
As the confluent hypergeometric function is defined as \cite{arfken}
\begin{equation}
    U(a, b, z)=\frac{1}{\Gamma(a)}\int_0^\infty \exp(-zt) t^{a-1}(1+t)^{b-a-1} dt. \label{eq:U}
\end{equation}
Comparing \eqref{eq:conv} and \eqref{eq:U} we identify
\begin{eqnarray}
a =  \frac{1}{q-1}-\frac{1}{2} & \text{ and } & b  =  \frac{1}{2}.
\end{eqnarray}
Thus, we can write the TMD as
\begin{equation}
    \frac{dN}{dp_T^2} \sim \frac{1}{\sqrt{\pi}}\Gamma \left( \frac{1}{q-1}  \right) U\left( \frac{1}{q-1}-\frac{1}{2}, \frac{1}{2}, \pi p_T^2 \frac{q-1}{2\sigma^2}  \right). \label{eq:U2}
\end{equation}
The function $U(a, b, z)$ has two well-known asymptotic behaviors. At low $z$,
\begin{equation}
    U(a,b,z)\approx \frac{\Gamma(1-b)}{\Gamma(1+a-b)} + \frac{\Gamma(b-1)}{\Gamma(a)} z^{1-b}.
\end{equation}
Thus
\begin{equation}
    \frac{dN}{dp_T^2} \sim \exp\left( -\frac{\sqrt{2\pi (q-1)}\Gamma\left( \frac{1}{q-1} \right) p_T}{\Gamma\left( \frac{1}{q-1}-\frac{1}{2} \right) \sigma}  \right).
\end{equation}
The effective or thermal temperature is
\begin{equation}
    T_{th}=\sigma \frac{\Gamma\left( \frac{1}{q-1}-\frac{1}{2} \right)}{\sqrt{2\pi (q-1)}\Gamma\left( \frac{1}{q-1} \right)}.
\end{equation}
On the other hand, at high $z$
\begin{equation}
    U(a, b, z) \approx z^{-a}.
\end{equation}
Thus, at high $p_T$ we obtain
\begin{equation}
    \frac{dN}{dp_T^2} \sim \frac{\Gamma\left( \frac{1}{q-1} \right)}{\sqrt{\pi}} \left(  \frac{\pi p_T^2 (q-1)}{2\sigma^2} \right)^{\frac{1}{2}-\frac{1}{q-1}},
\end{equation}
and the hard temperature is
\begin{equation}
    T_H= \sigma \sqrt{\frac{2}{\pi(q-1)}} \left( \frac{\sqrt{\pi}}{\Gamma\left( \frac{1}{q-1} \right)}  \right)^{\frac{q-1}{q-3}}. 
\end{equation}
The ratio between both temperatures, $T_{th}$ and $T_H$, is
\begin{equation}
    \frac{T_H}{T_{th}}=2 \left( \frac{\sqrt{\pi}}{\Gamma\left( \frac{1}{q-1} \right)}  \right)^{\frac{q-1}{q-3}} \frac{\Gamma\left( \frac{1}{q-1} \right)}{\Gamma\left( \frac{1}{q-1}-\frac{1}{2} \right)}. \label{eq:ratio}
\end{equation}
In Fig.~\ref{fig:ratio} we plot the ratio $T_H/T_{th}$ for reasonable values of $q$ such that Eq.~\eqref{eq:ratio} takes values between 2 and 5.5.

\section{Discussion}

\begin{figure*}[h]
\centering
\includegraphics[scale=1]{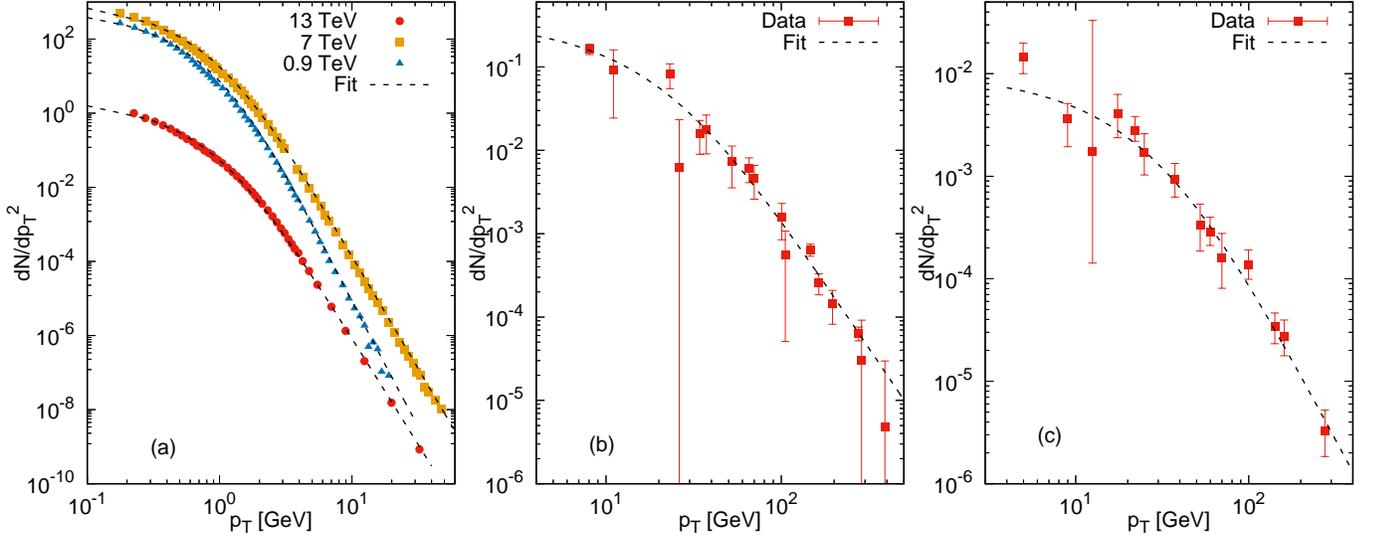}
\caption{(a) Charged particle TMD in pp collisions in the centrality class 0\%-5\% at $\sqrt{s}=$0.9 TeV (triangles), 7 TeV (squares), and 13 TeV (circles).
(b) Differential cross section of the Higgs boson production from the process H$\to\gamma\gamma$ decay in pp collisions at $\sqrt{s}=$13 TeV.
(c) Differential cross section of the Higgs boson production from the process H$\to4l$ decay in pp collisions at $\sqrt{s}=$13 TeV.
Data extracted from Refs.~\cite{feal2, baker, datapp}}.
\label{fig:tmd}
\end{figure*}

\begin{table*}[h]
\caption{Values of $T_{th}$, $T_H$, and $q$ for the processes analyzed.  \label{tab:tab}}
\begin{center}
\begin{tabular}{c c c c c}
\hline
Process &  $T_{th}$ [GeV] &  $T_{H}$ [GeV] & $q$ & $\chi^2$/NDF\\
\hline
pp at $\sqrt{s}$=0.9 TeV $\rightarrow$ charged hadrons & 0.189$\pm$0.003 & 0.90$\pm$0.08 & 1.23$\pm$0.02 & 39/51  \\
pp at $\sqrt{s}$=7 TeV $\rightarrow$ charged hadrons & 0.197$\pm$0.002 & 0.737$\pm$0.009 & 1.282$\pm$0.002 & 39/59\\
pp at $\sqrt{s}$=13 TeV $\rightarrow$ charged hadrons & 0.211$\pm$0.005 & 0.75$\pm$0.03 & 1.296$\pm$0.009 & 12.5/41\\
pp at $\sqrt{s}$=13 TeV $\rightarrow$ H $\rightarrow$ $\gamma\gamma$ & 9.0$\pm$0.1 & 17.5$\pm$0.2 & 1.485$\pm$0.003 & 13.3/14 \\
pp at $\sqrt{s}$=13 TeV $\rightarrow$ H $\rightarrow$ 4$l$ & 12.1$\pm$0.1 & 23.8$\pm$0.2 & 1.480$\pm$0.003 & 5.9/11 \\
\hline
\end{tabular} 
\end{center}
\end{table*}

The obtained universal function in Eq.~\eqref{eq:U2} has the right behavior at low and high transverse momentum that was used in Refs. \cite{baker, feal, feal2}, therefore we reproduce most of their results. 
In Fig.~\ref{fig:tmd} we show the fits obtained for some processes and their values $T_{th}$, $T_H$, and $q$ are summarized in Table~\ref{tab:tab}. 
Our results of the thermal and hard temperatures are in agreement with the reported in \cite{baker, feal, Bellwied_2018, feal2} for the TMD for the production of charged particles in pp collisions at $\sqrt{s}$=0.9 TeV, 7 TeV, and 13 TeV (see Fig.~\ref{fig:tmd} (a)).
Notice that $q$ increases slowly with energy (and/or centrality), implying that the thermal scale as well increases. It follows the same trend shown in the aforementioned references. On the other hand, we cannot conclude a particular trend for the hard scale since it can be computed as the multiplication of the thermal scale by the l.h.s of Eq.~\eqref{eq:ratio}, where the former is a slowly increasing function of $q$ while the latter is a decreasing function of $q$.

Other processes studied are H$\to\gamma\gamma$ and H$\to 4l$ (see Fig.~\ref{fig:tmd} (b) and Fig.~\ref{fig:tmd} (c), respectively). In these cases our results for both scales differ for the values reported in Ref.~\cite{baker}, because the scale parameter $\sigma$ of the confluent hypergeometric function is sensitive to the selected range to perform the fits. In particular, the low $p_T$ values of the TMD for the H$\to\gamma\gamma$ and H$\to 4l$ processes are approximately two order of magnitude larger than the cases of pp collisions.

The use of the q-Gaussian distribution for the string tension fluctuations has allowed the obtention of a universal analytical function that covers all the spectrum.
The extension of the Gaussian shape to the q-Gaussian in some sense is natural because is the most economical way of doing it introducing a single parameter that is related to the hard scale of the spectrum. 
In this way, the hard part of the collisions and the  string tension fluctuations are related.
This is in agreement with the observation that a hard collision between two partons of the inicial state of the colliding hadrons produces a rapid quench of the entangled partonic state, and thus the effective thermal temperature distribution of the produces particles \cite{Calabrese_2016, BERGES2018442, kharzeev, armesto2019, gotsman, tu, Ramos}.
Moreover, the parameter $q$ marks the departure of the thermal behavior and thus the degree of hardness of the process under analysis. Note that the limit $q\to1$ recovers the exponential decay of the TMD at low $p_T$.

In the same way that the Gaussian distribution obeys the minimum uncertainty Heisenberg principle, the q-Gaussian distribution obeys the generalized uncertainty minimum principle (GUP) \cite{jizba}. GUP was introduced in quantum gravity due to the existence of a minimal length scale of the order of the Planck length \cite{kempf, magueijo}.
In the usual uncertainty principle, the minimal measurable length $\Delta x$ can be made arbitrarily small by allowing $\Delta p$ to grow correspondingly. This is not possible if there is a minimal length scale and thus results in a generalized uncertainty principle. As a consequence, it modifies the commutation relation between $x$ and $p$. Now it reads \cite{shababi, luciano}
\begin{equation}
    [x,p]=i\hbar(1+\beta p^2),
\end{equation}
and
\begin{equation}
    \Delta x \Delta p \geq \frac{\hbar}{2}\left( 1+ \Delta p^2 \beta  \right). \label{eq:GUP}
\end{equation}
This equation implies a lower bound for $\Delta x$ that can be obtained by saturating the above inequality, to give
\begin{equation}
    \Delta x_{min}=\hbar \sqrt{\beta}=\sqrt{\beta_0}l_p,
\end{equation}
where $l_p$ is the Planck length. The value of $\beta_0$ must not be far from one to ensure that $\Delta x_{min}$ is of the order of the Planck length.

The existence of a minimal length is not restricted to quantum gravity, but occurs in the study of different systems \cite{shababi}.
For instance, a gas of particles which have a size of the order of the atom radius.
In our case, a high $p_T$ particle, probes a spatial size of the order of $1/p_T$.
An experiment at a given energy have a limited $p_T$, therefore have an intrinsic minimum length. In this case the value of $\beta$ in Eq.~\eqref{eq:GUP} is proportional to $q-1$ which marks the degree of hardness of the collision and thus the limited length to probe.

The minimum uncertainty of the GUP in Eq.~\eqref{eq:GUP} is satisfied by the q-Gaussian distribution \cite{jizba}. In this way plays the same role in GUP than the Gaussian function in the Heisenberg uncertainty principle.

The thermal and hard scales depend on $q$ and $\sigma$. Both parameters depend on the energy and multiplicity of the collisions. In some models this dependence is determined.
For instance, in the color string percolation model (CSPM) \cite{BRAUN20151, string}, the width $\sigma$ of the string tension distribution is related to the string density.
In this model, the tension of a cluster of overlapped strings is proportional to the color field which results from sum the individual color field of each strings.
Due to the randomness in color space of the direction of individual color fields, the intensity of the resulting color  field is only the squared root of the number of overlapped strings of the cluster, which is different to the number of strings times the intensity of one string.
In this way, the multiplicity and the average of $p_T^2$ are given by
\begin{eqnarray}
\mu_n=\sqrt{\frac{nS_n}{S_1}}\mu_1 & \text{and} & \langle p_T^2 \rangle_n = \sqrt{\frac{nS_1}{S_n}} \langle p_T^2 \rangle_1, \label{eq:mun}
\end{eqnarray}
where $S_n$ and $S_1$ are the area covered by a cluster of $n$ strings and a single string, respectively. $\mu_1$ and $\langle p_T^2 \rangle_1$ are the multiplicity and the mean squared $p_T$ of the produced particles in the fragmentation of one string.
Taking into account the fluctuations of the string number at fixed string density, the relations in Eq.~\eqref{eq:mun} transform into
\begin{eqnarray}
\mu=nF(\xi) \mu_1 & \text{and} & \langle p_T^2 \rangle = \frac{ \langle p_T^2 \rangle_1}{F(\xi)}, \label{eq:mu}
\end{eqnarray}
with
\begin{equation}
    F(\xi)=\sqrt{\frac{1-\exp(-\xi)}{\xi}}
\end{equation}
being the color suppression function at the thermodynamic limit and $\xi=nS_1/S$ is the string density \cite{braun2000, braunprl}.
Thus the parameter $\sigma$ and both scales, $T_{th}$ and $T_H$, can be related to $\langle p_T^2 \rangle_1/F(\xi)$.
In this way it is obtained a universal behavior of the fluctuation for all type of collisions depending only on the string density. On the other hand, in the CSPM, $q-1$ is related to the normalized multiplicity fluctuations.

\section{Conclusions}
The introduction of a $q$-Gaussian distribution to describe the string tension fluctuations allows to obtain the right thermal behavior at low $p_T$ from the Gaussian shape transverse momentum distributions of the produced particles coming from the fragmentation of a string, via the Schwinger mechanism, as it is done by most of the string models.
Also, it gives rise to the power like behavior of the high $p_T$ part of the spectrum.
The low and high $p_T$ behavior of the TMD are encoded in a single analytical function, the confluent hypergeometric function, which has two parameters, $\sigma$ and $q$. 
In particular, $\sigma$ is the width of the $q$-Gaussian distribution. In the limit $q\to1$, the $q$-Gaussian becomes into the Gaussian distribution.
The parameter $q$ determines the departure of the thermal collision as well as the power like part of the spectrum. In the CSPM, $\sigma$ is given by a function of the string density. Given an energy or centrality, it is known the string density and thus this parameter. 
Moreover, the ratio between the hard and thermal scales is a function that only depends on $q$.

It is pointed out that the $q$-Gaussian satisfies the equality of the Generalized Uncertainty Principle. The correction therm of the GUP to the Heisenberg uncertainty principle is proportional to the existence of a minimum length. In a collision, this term is set by $q-1$ because establishes the minimum length that can be probed in a high $p_T$ collision.

\begin{acknowledgments}
We thank X. Feal and R. A. Vázquez who participated in the early stages of this study.
C. P. thanks the grant Maria de Maeztu Unit of Excelence under the project MDM-2016 0682 of Ministry of Science and Innovation of Spain. This work has been funded by the projects PID2020-119632GB-100 of the Spanish Research Agency, Centro Singular de Galicia 2019-2022 of Xunta de Galicia and the ERDF of the European Union.
\end{acknowledgments}

\bibliography{bib}

\begin{thebibliography}{29}%
\makeatletter
\providecommand \@ifxundefined [1]{%
 \@ifx{#1\undefined}
}%
\providecommand \@ifnum [1]{%
 \ifnum #1\expandafter \@firstoftwo
 \else \expandafter \@secondoftwo
 \fi
}%
\providecommand \@ifx [1]{%
 \ifx #1\expandafter \@firstoftwo
 \else \expandafter \@secondoftwo
 \fi
}%
\providecommand \natexlab [1]{#1}%
\providecommand \enquote  [1]{``#1''}%
\providecommand \bibnamefont  [1]{#1}%
\providecommand \bibfnamefont [1]{#1}%
\providecommand \citenamefont [1]{#1}%
\providecommand \href@noop [0]{\@secondoftwo}%
\providecommand \href [0]{\begingroup \@sanitize@url \@href}%
\providecommand \@href[1]{\@@startlink{#1}\@@href}%
\providecommand \@@href[1]{\endgroup#1\@@endlink}%
\providecommand \@sanitize@url [0]{\catcode `\\12\catcode `\$12\catcode
  `\&12\catcode `\#12\catcode `\^12\catcode `\_12\catcode `\%12\relax}%
\providecommand \@@startlink[1]{}%
\providecommand \@@endlink[0]{}%
\providecommand \url  [0]{\begingroup\@sanitize@url \@url }%
\providecommand \@url [1]{\endgroup\@href {#1}{\urlprefix }}%
\providecommand \urlprefix  [0]{URL }%
\providecommand \Eprint [0]{\href }%
\providecommand \doibase [0]{http://dx.doi.org/}%
\providecommand \selectlanguage [0]{\@gobble}%
\providecommand \bibinfo  [0]{\@secondoftwo}%
\providecommand \bibfield  [0]{\@secondoftwo}%
\providecommand \translation [1]{[#1]}%
\providecommand \BibitemOpen [0]{}%
\providecommand \bibitemStop [0]{}%
\providecommand \bibitemNoStop [0]{.\EOS\space}%
\providecommand \EOS [0]{\spacefactor3000\relax}%
\providecommand \BibitemShut  [1]{\csname bibitem#1\endcsname}%
\let\auto@bib@innerbib\@empty
\bibitem [{\citenamefont {Bylinkin}\ and\ \citenamefont
  {Rostovtsev}(2014)}]{BYLINKIN201465}%
  \BibitemOpen
  \bibfield  {author} {\bibinfo {author} {\bibfnamefont {A.~A.}\ \bibnamefont
  {Bylinkin}}\ and\ \bibinfo {author} {\bibfnamefont {A.~A.}\ \bibnamefont
  {Rostovtsev}},\ }\href {\doibase
  https://doi.org/10.1016/j.nuclphysb.2014.09.010} {\bibfield  {journal}
  {\bibinfo  {journal} {Nucl. Phys. B}\ }\textbf {\bibinfo {volume} {888}},\
  \bibinfo {pages} {65} (\bibinfo {year} {2014})}\BibitemShut {NoStop}%
\bibitem [{\citenamefont {Bylinkin}\ \emph {et~al.}(2014)\citenamefont
  {Bylinkin}, \citenamefont {Kharzeev},\ and\ \citenamefont
  {Rostovtsev}}]{bylinkin}%
  \BibitemOpen
  \bibfield  {author} {\bibinfo {author} {\bibfnamefont {A.~A.}\ \bibnamefont
  {Bylinkin}}, \bibinfo {author} {\bibfnamefont {D.~E.}\ \bibnamefont
  {Kharzeev}}, \ and\ \bibinfo {author} {\bibfnamefont {A.~A.}\ \bibnamefont
  {Rostovtsev}},\ }\href {\doibase 10.1142/S0218301314500839} {\bibfield
  {journal} {\bibinfo  {journal} {Int. J. Mod. Phys. E}\ }\textbf {\bibinfo
  {volume} {23}},\ \bibinfo {pages} {1450083} (\bibinfo {year}
  {2014})}\BibitemShut {NoStop}%
\bibitem [{\citenamefont {Bialas}(1999)}]{BIALAS1999301}%
  \BibitemOpen
  \bibfield  {author} {\bibinfo {author} {\bibfnamefont {A.}~\bibnamefont
  {Bialas}},\ }\href {\doibase https://doi.org/10.1016/S0370-2693(99)01159-4}
  {\bibfield  {journal} {\bibinfo  {journal} {Phys. Lett. B}\ }\textbf
  {\bibinfo {volume} {466}},\ \bibinfo {pages} {301} (\bibinfo {year}
  {1999})}\BibitemShut {NoStop}%
\bibitem [{\citenamefont {{Dias de Deus}}\ and\ \citenamefont
  {Pajares}(2006)}]{DIASDEDEUS2006455}%
  \BibitemOpen
  \bibfield  {author} {\bibinfo {author} {\bibfnamefont {J.}~\bibnamefont
  {{Dias de Deus}}}\ and\ \bibinfo {author} {\bibfnamefont {C.}~\bibnamefont
  {Pajares}},\ }\href {\doibase https://doi.org/10.1016/j.physletb.2006.10.018}
  {\bibfield  {journal} {\bibinfo  {journal} {Phys. Lett. B}\ }\textbf
  {\bibinfo {volume} {642}},\ \bibinfo {pages} {455} (\bibinfo {year}
  {2006})}\BibitemShut {NoStop}%
\bibitem [{\citenamefont {Kharzeev}\ and\ \citenamefont
  {Tuchin}(2005)}]{KHARZEEV2005316}%
  \BibitemOpen
  \bibfield  {author} {\bibinfo {author} {\bibfnamefont {D.}~\bibnamefont
  {Kharzeev}}\ and\ \bibinfo {author} {\bibfnamefont {K.}~\bibnamefont
  {Tuchin}},\ }\href {\doibase https://doi.org/10.1016/j.nuclphysa.2005.03.001}
  {\bibfield  {journal} {\bibinfo  {journal} {Nucl. Phys. A}\ }\textbf
  {\bibinfo {volume} {753}},\ \bibinfo {pages} {316} (\bibinfo {year}
  {2005})}\BibitemShut {NoStop}%
\bibitem [{\citenamefont {Costin}\ and\ \citenamefont
  {Dunne}(2020)}]{COSTIN2020135627}%
  \BibitemOpen
  \bibfield  {author} {\bibinfo {author} {\bibfnamefont {O.}~\bibnamefont
  {Costin}}\ and\ \bibinfo {author} {\bibfnamefont {G.~V.}\ \bibnamefont
  {Dunne}},\ }\href {\doibase https://doi.org/10.1016/j.physletb.2020.135627}
  {\bibfield  {journal} {\bibinfo  {journal} {Phys. Lett. B}\ }\textbf
  {\bibinfo {volume} {808}},\ \bibinfo {pages} {135627} (\bibinfo {year}
  {2020})}\BibitemShut {NoStop}%
\bibitem [{\citenamefont {Calabrese}\ and\ \citenamefont
  {Cardy}(2016)}]{Calabrese_2016}%
  \BibitemOpen
  \bibfield  {author} {\bibinfo {author} {\bibfnamefont {P.}~\bibnamefont
  {Calabrese}}\ and\ \bibinfo {author} {\bibfnamefont {J.}~\bibnamefont
  {Cardy}},\ }\href {\doibase 10.1088/1742-5468/2016/06/064003} {\bibfield
  {journal} {\bibinfo  {journal} {J. Stat. Mech.- Theory E}\ }\textbf {\bibinfo
  {volume} {2016}},\ \bibinfo {pages} {064003} (\bibinfo {year}
  {2016})}\BibitemShut {NoStop}%
\bibitem [{\citenamefont {Berges}\ \emph {et~al.}(2018)\citenamefont {Berges},
  \citenamefont {Floerchinger},\ and\ \citenamefont
  {Venugopalan}}]{BERGES2018442}%
  \BibitemOpen
  \bibfield  {author} {\bibinfo {author} {\bibfnamefont {J.}~\bibnamefont
  {Berges}}, \bibinfo {author} {\bibfnamefont {S.}~\bibnamefont
  {Floerchinger}}, \ and\ \bibinfo {author} {\bibfnamefont {R.}~\bibnamefont
  {Venugopalan}},\ }\href {\doibase
  https://doi.org/10.1016/j.physletb.2018.01.068} {\bibfield  {journal}
  {\bibinfo  {journal} {Phys. Lett. B}\ }\textbf {\bibinfo {volume} {778}},\
  \bibinfo {pages} {442} (\bibinfo {year} {2018})}\BibitemShut {NoStop}%
\bibitem [{\citenamefont {Kharzeev}\ and\ \citenamefont
  {Levin}(2017)}]{kharzeev}%
  \BibitemOpen
  \bibfield  {author} {\bibinfo {author} {\bibfnamefont {D.~E.}\ \bibnamefont
  {Kharzeev}}\ and\ \bibinfo {author} {\bibfnamefont {E.~M.}\ \bibnamefont
  {Levin}},\ }\href {\doibase 10.1103/PhysRevD.95.114008} {\bibfield  {journal}
  {\bibinfo  {journal} {Phys. Rev. D}\ }\textbf {\bibinfo {volume} {95}},\
  \bibinfo {pages} {114008} (\bibinfo {year} {2017})}\BibitemShut {NoStop}%
\bibitem [{\citenamefont {Armesto}\ \emph {et~al.}(2019)\citenamefont
  {Armesto}, \citenamefont {Domínguez}, \citenamefont {Kovner}, \citenamefont
  {Lublinsky},\ and\ \citenamefont {Skokov}}]{armesto2019}%
  \BibitemOpen
  \bibfield  {author} {\bibinfo {author} {\bibfnamefont {N.}~\bibnamefont
  {Armesto}}, \bibinfo {author} {\bibfnamefont {F.}~\bibnamefont {Domínguez}},
  \bibinfo {author} {\bibfnamefont {A.}~\bibnamefont {Kovner}}, \bibinfo
  {author} {\bibfnamefont {M.}~\bibnamefont {Lublinsky}}, \ and\ \bibinfo
  {author} {\bibfnamefont {V.~V.}\ \bibnamefont {Skokov}},\ }\href {\doibase
  doi.org/10.1007/JHEP05(2019)025} {\bibfield  {journal} {\bibinfo  {journal}
  {J. High Energy Phys.}\ }\textbf {\bibinfo {volume} {2019}},\ \bibinfo
  {pages} {25} (\bibinfo {year} {2019})}\BibitemShut {NoStop}%
\bibitem [{\citenamefont {Gotsman}\ and\ \citenamefont
  {Levin}(2020)}]{gotsman}%
  \BibitemOpen
  \bibfield  {author} {\bibinfo {author} {\bibfnamefont {E.}~\bibnamefont
  {Gotsman}}\ and\ \bibinfo {author} {\bibfnamefont {E.}~\bibnamefont
  {Levin}},\ }\href {\doibase 10.1103/PhysRevD.102.074008} {\bibfield
  {journal} {\bibinfo  {journal} {Phys. Rev. D}\ }\textbf {\bibinfo {volume}
  {102}},\ \bibinfo {pages} {074008} (\bibinfo {year} {2020})}\BibitemShut
  {NoStop}%
\bibitem [{\citenamefont {Tu}\ \emph {et~al.}(2020)\citenamefont {Tu},
  \citenamefont {Kharzeev},\ and\ \citenamefont {Ullrich}}]{tu}%
  \BibitemOpen
  \bibfield  {author} {\bibinfo {author} {\bibfnamefont {Z.}~\bibnamefont
  {Tu}}, \bibinfo {author} {\bibfnamefont {D.~E.}\ \bibnamefont {Kharzeev}}, \
  and\ \bibinfo {author} {\bibfnamefont {T.}~\bibnamefont {Ullrich}},\ }\href
  {\doibase 10.1103/PhysRevLett.124.062001} {\bibfield  {journal} {\bibinfo
  {journal} {Phys. Rev. Lett.}\ }\textbf {\bibinfo {volume} {124}},\ \bibinfo
  {pages} {062001} (\bibinfo {year} {2020})}\BibitemShut {NoStop}%
\bibitem [{\citenamefont {Ramos}\ and\ \citenamefont {Machado}(2020)}]{Ramos}%
  \BibitemOpen
  \bibfield  {author} {\bibinfo {author} {\bibfnamefont {G.~S.}\ \bibnamefont
  {Ramos}}\ and\ \bibinfo {author} {\bibfnamefont {M.~V.~T.}\ \bibnamefont
  {Machado}},\ }\href {\doibase 10.1103/PhysRevD.101.074040} {\bibfield
  {journal} {\bibinfo  {journal} {Phys. Rev. D}\ }\textbf {\bibinfo {volume}
  {101}},\ \bibinfo {pages} {074040} (\bibinfo {year} {2020})}\BibitemShut
  {NoStop}%
\bibitem [{\citenamefont {Baker}\ and\ \citenamefont {Kharzeev}(2018)}]{baker}%
  \BibitemOpen
  \bibfield  {author} {\bibinfo {author} {\bibfnamefont {O.~K.}\ \bibnamefont
  {Baker}}\ and\ \bibinfo {author} {\bibfnamefont {D.~E.}\ \bibnamefont
  {Kharzeev}},\ }\href {\doibase 10.1103/PhysRevD.98.054007} {\bibfield
  {journal} {\bibinfo  {journal} {Phys. Rev. D}\ }\textbf {\bibinfo {volume}
  {98}},\ \bibinfo {pages} {054007} (\bibinfo {year} {2018})}\BibitemShut
  {NoStop}%
\bibitem [{\citenamefont {Feal}\ \emph {et~al.}(2019)\citenamefont {Feal},
  \citenamefont {Pajares},\ and\ \citenamefont {Vazquez}}]{feal}%
  \BibitemOpen
  \bibfield  {author} {\bibinfo {author} {\bibfnamefont {X.}~\bibnamefont
  {Feal}}, \bibinfo {author} {\bibfnamefont {C.}~\bibnamefont {Pajares}}, \
  and\ \bibinfo {author} {\bibfnamefont {R.~A.}\ \bibnamefont {Vazquez}},\
  }\href {\doibase 10.1103/PhysRevC.99.015205} {\bibfield  {journal} {\bibinfo
  {journal} {Phys. Rev. C}\ }\textbf {\bibinfo {volume} {99}},\ \bibinfo
  {pages} {015205} (\bibinfo {year} {2019})}\BibitemShut {NoStop}%
\bibitem [{\citenamefont {Bellwied}(2018)}]{Bellwied_2018}%
  \BibitemOpen
  \bibfield  {author} {\bibinfo {author} {\bibfnamefont {R.}~\bibnamefont
  {Bellwied}},\ }\href {\doibase 10.1088/1742-6596/1070/1/012001} {\bibfield
  {journal} {\bibinfo  {journal} {J. Phys. Conf. Ser.}\ }\textbf {\bibinfo
  {volume} {1070}},\ \bibinfo {pages} {012001} (\bibinfo {year}
  {2018})}\BibitemShut {NoStop}%
\bibitem [{\citenamefont {Feal}\ \emph {et~al.}(2021)\citenamefont {Feal},
  \citenamefont {Pajares},\ and\ \citenamefont {Vazquez}}]{feal2}%
  \BibitemOpen
  \bibfield  {author} {\bibinfo {author} {\bibfnamefont {X.}~\bibnamefont
  {Feal}}, \bibinfo {author} {\bibfnamefont {C.}~\bibnamefont {Pajares}}, \
  and\ \bibinfo {author} {\bibfnamefont {R.~A.}\ \bibnamefont {Vazquez}},\
  }\href {\doibase 10.1103/PhysRevC.104.044904} {\bibfield  {journal} {\bibinfo
   {journal} {Phys. Rev. C}\ }\textbf {\bibinfo {volume} {104}},\ \bibinfo
  {pages} {044904} (\bibinfo {year} {2021})}\BibitemShut {NoStop}%
\bibitem [{\citenamefont {Budini}(2015)}]{budini}%
  \BibitemOpen
  \bibfield  {author} {\bibinfo {author} {\bibfnamefont {A.~A.}\ \bibnamefont
  {Budini}},\ }\href {\doibase 10.1103/PhysRevE.91.052113} {\bibfield
  {journal} {\bibinfo  {journal} {Phys. Rev. E}\ }\textbf {\bibinfo {volume}
  {91}},\ \bibinfo {pages} {052113} (\bibinfo {year} {2015})}\BibitemShut
  {NoStop}%
\bibitem [{\citenamefont {Arfken}(1995)}]{arfken}%
  \BibitemOpen
  \bibfield  {author} {\bibinfo {author} {\bibfnamefont {G.~B.}\ \bibnamefont
  {Arfken}},\ }\href@noop {} {\emph {\bibinfo {title} {{Mathematical Methods
  for Physicists}}}}\ (\bibinfo  {publisher} {Academic Press},\ \bibinfo
  {address} {California},\ \bibinfo {year} {1995})\ Chap.~\bibinfo {chapter}
  {13}, pp.\ \bibinfo {pages} {802--803}\BibitemShut {NoStop}%
\bibitem [{\citenamefont {{The ALICE Collaboration}}(2013)}]{datapp}%
  \BibitemOpen
  \bibfield  {author} {\bibinfo {author} {\bibnamefont {{The ALICE
  Collaboration}}},\ }\href@noop {} {\bibfield  {journal} {\bibinfo  {journal}
  {Eur. Phys. J. C}\ }\textbf {\bibinfo {volume} {73}},\ \bibinfo {pages}
  {2662} (\bibinfo {year} {2013})}\BibitemShut {NoStop}%
\bibitem [{\citenamefont {Jizba}\ \emph {et~al.}(2022)\citenamefont {Jizba},
  \citenamefont {Lambiase}, \citenamefont {Luciano},\ and\ \citenamefont
  {Petruzziello}}]{jizba}%
  \BibitemOpen
  \bibfield  {author} {\bibinfo {author} {\bibfnamefont {P.}~\bibnamefont
  {Jizba}}, \bibinfo {author} {\bibfnamefont {G.}~\bibnamefont {Lambiase}},
  \bibinfo {author} {\bibfnamefont {G.~G.}\ \bibnamefont {Luciano}}, \ and\
  \bibinfo {author} {\bibfnamefont {L.}~\bibnamefont {Petruzziello}},\ }\href
  {\doibase 10.1103/PhysRevD.105.L121501} {\bibfield  {journal} {\bibinfo
  {journal} {Phys. Rev. D}\ }\textbf {\bibinfo {volume} {105}},\ \bibinfo
  {pages} {L121501} (\bibinfo {year} {2022})}\BibitemShut {NoStop}%
\bibitem [{\citenamefont {Kempf}\ \emph {et~al.}(1995)\citenamefont {Kempf},
  \citenamefont {Mangano},\ and\ \citenamefont {Mann}}]{kempf}%
  \BibitemOpen
  \bibfield  {author} {\bibinfo {author} {\bibfnamefont {A.}~\bibnamefont
  {Kempf}}, \bibinfo {author} {\bibfnamefont {G.}~\bibnamefont {Mangano}}, \
  and\ \bibinfo {author} {\bibfnamefont {R.~B.}\ \bibnamefont {Mann}},\ }\href
  {\doibase 10.1103/PhysRevD.52.1108} {\bibfield  {journal} {\bibinfo
  {journal} {Phys. Rev. D}\ }\textbf {\bibinfo {volume} {52}},\ \bibinfo
  {pages} {1108} (\bibinfo {year} {1995})}\BibitemShut {NoStop}%
\bibitem [{\citenamefont {Magueijo}\ and\ \citenamefont
  {Smolin}(2002)}]{magueijo}%
  \BibitemOpen
  \bibfield  {author} {\bibinfo {author} {\bibfnamefont {J.~a.}\ \bibnamefont
  {Magueijo}}\ and\ \bibinfo {author} {\bibfnamefont {L.}~\bibnamefont
  {Smolin}},\ }\href {\doibase 10.1103/PhysRevLett.88.190403} {\bibfield
  {journal} {\bibinfo  {journal} {Phys. Rev. Lett.}\ }\textbf {\bibinfo
  {volume} {88}},\ \bibinfo {pages} {190403} (\bibinfo {year}
  {2002})}\BibitemShut {NoStop}%
\bibitem [{\citenamefont {Shababi}\ and\ \citenamefont
  {Ourabah}(2020)}]{shababi}%
  \BibitemOpen
  \bibfield  {author} {\bibinfo {author} {\bibfnamefont {H.}~\bibnamefont
  {Shababi}}\ and\ \bibinfo {author} {\bibfnamefont {K.}~\bibnamefont
  {Ourabah}},\ }\href {\doibase 10.1140/epjp/s13360-020-00726-9} {\bibfield
  {journal} {\bibinfo  {journal} {Eur. Phys. J. Plus}\ }\textbf {\bibinfo
  {volume} {135}},\ \bibinfo {pages} {697} (\bibinfo {year}
  {2020})}\BibitemShut {NoStop}%
\bibitem [{\citenamefont {Luciano}(2021)}]{luciano}%
  \BibitemOpen
  \bibfield  {author} {\bibinfo {author} {\bibfnamefont {G.~G.}\ \bibnamefont
  {Luciano}},\ }\href {\doibase 10.1140/epjc/s10052-021-09486-x} {\bibfield
  {journal} {\bibinfo  {journal} {Eur. Phys. J. C}\ }\textbf {\bibinfo {volume}
  {81}},\ \bibinfo {pages} {672} (\bibinfo {year} {2021})}\BibitemShut
  {NoStop}%
\bibitem [{\citenamefont {Braun}\ \emph {et~al.}(2015)\citenamefont {Braun},
  \citenamefont {{Dias de Deus}}, \citenamefont {Hirsch}, \citenamefont
  {Pajares}, \citenamefont {Scharenberg},\ and\ \citenamefont
  {Srivastava}}]{BRAUN20151}%
  \BibitemOpen
  \bibfield  {author} {\bibinfo {author} {\bibfnamefont {M.}~\bibnamefont
  {Braun}}, \bibinfo {author} {\bibfnamefont {J.}~\bibnamefont {{Dias de
  Deus}}}, \bibinfo {author} {\bibfnamefont {A.}~\bibnamefont {Hirsch}},
  \bibinfo {author} {\bibfnamefont {C.}~\bibnamefont {Pajares}}, \bibinfo
  {author} {\bibfnamefont {R.}~\bibnamefont {Scharenberg}}, \ and\ \bibinfo
  {author} {\bibfnamefont {B.}~\bibnamefont {Srivastava}},\ }\href {\doibase
  https://doi.org/10.1016/j.physrep.2015.09.003} {\bibfield  {journal}
  {\bibinfo  {journal} {Phys. Rep.}\ }\textbf {\bibinfo {volume} {599}},\
  \bibinfo {pages} {1} (\bibinfo {year} {2015})}\BibitemShut {NoStop}%
\bibitem [{\citenamefont {Bautista}\ \emph {et~al.}(2019)\citenamefont
  {Bautista}, \citenamefont {Pajares},\ and\ \citenamefont
  {Ram{\'\i}rez}}]{string}%
  \BibitemOpen
  \bibfield  {author} {\bibinfo {author} {\bibfnamefont {I.}~\bibnamefont
  {Bautista}}, \bibinfo {author} {\bibfnamefont {C.}~\bibnamefont {Pajares}}, \
  and\ \bibinfo {author} {\bibfnamefont {J.~E.}\ \bibnamefont {Ram{\'\i}rez}},\
  }\href {\doibase https://doi.org/10.31349/RevMexFis.65.197} {\bibfield
  {journal} {\bibinfo  {journal} {Rev. Mex. Fis.}\ }\textbf {\bibinfo {volume}
  {65}},\ \bibinfo {pages} {197} (\bibinfo {year} {2019})}\BibitemShut
  {NoStop}%
\bibitem [{\citenamefont {Braun}\ and\ \citenamefont
  {Pajares}(2000{\natexlab{a}})}]{braun2000}%
  \BibitemOpen
  \bibfield  {author} {\bibinfo {author} {\bibfnamefont {M.~A.}\ \bibnamefont
  {Braun}}\ and\ \bibinfo {author} {\bibfnamefont {C.}~\bibnamefont
  {Pajares}},\ }\href {\doibase 10.1007/s100520050027} {\bibfield  {journal}
  {\bibinfo  {journal} {Eur. Phys. J. C}\ }\textbf {\bibinfo {volume} {16}},\
  \bibinfo {pages} {349} (\bibinfo {year} {2000}{\natexlab{a}})}\BibitemShut
  {NoStop}%
\bibitem [{\citenamefont {Braun}\ and\ \citenamefont
  {Pajares}(2000{\natexlab{b}})}]{braunprl}%
  \BibitemOpen
  \bibfield  {author} {\bibinfo {author} {\bibfnamefont {M.~A.}\ \bibnamefont
  {Braun}}\ and\ \bibinfo {author} {\bibfnamefont {C.}~\bibnamefont
  {Pajares}},\ }\href {\doibase 10.1103/PhysRevLett.85.4864} {\bibfield
  {journal} {\bibinfo  {journal} {Phys. Rev. Lett.}\ }\textbf {\bibinfo
  {volume} {85}},\ \bibinfo {pages} {4864} (\bibinfo {year}
  {2000}{\natexlab{b}})}\BibitemShut {NoStop}%
\end{thebibliography}%

\end{document}